\documentstyle{article}
\newfont{\footsc}{cmcsc10 at 8truept}
\newfont{\footbf}{cmbx10 at 8truept}
\newfont{\footrm}{cmr10 at 10truept}

\begin{document}

\centerline{\Large \bf On correlation functions
of characteristic polynomials}
\centerline{\Large \bf for chiral Gaussian Unitary Ensemble}
\vskip 0.5cm
\centerline{\large \bf Yan V Fyodorov\footnote{
on leave from Petersburg Nuclear Physics Institute, Gatchina,
Russia}
and Eugene Strahov}

\centerline{Department of Mathematical Sciences, Brunel University}
\centerline{Uxbridge, UB8 3PH, United Kingdom}
\centerline{\small \texttt{Yan.Fyodorov@brunel.ac.uk}}
\centerline{\small \texttt{Eugene.Strahov@brunel.ac.uk}}

\vskip0.3cm

\begin{abstract}
We calculate a
 general spectral correlation function  of
 products and ratios of characteristic polynomials for a $N\times N$
random matrix taken from the chiral Gaussian Unitary Ensemble
(chGUE). Our derivation is based upon finding a  Itzykson-Zuber
type integral for matrices from the non-compact manifold
${\sf{Gl(n,{\mathcal{C}})/U(1)\times ...\times U(1)}}$ (matrix
Macdonald function). The correlation function is shown to be
always represented in a determinant form generalising the known
expressions for only positive moments. Finally, we present the
asymptotic formula for the correlation function in the large
matrix size limit.

\end{abstract}

\section{Introduction}

A class of random matrices that has attracted a considerable
attention recently
\cite{SN,VeZa,For,TrWi,AST,JSV1,JSV2,GW,JNZ,WGW,DN,BH,Ak,DV} is
the so-called {\it chiral} GUE, also known as the Laguerre
ensemble. The corresponding matrices are of the form
$\hat{D}=\left(\begin{array}{cc}{\bf 0}&\hat{J}\\
\hat{J}^{\dagger}& {\bf 0} \end{array}\right)$, where $\hat{J}$
stands for a complex matrix, with $\hat{J}^{\dagger}$ being its
Hermitian conjugate.

The off-diagonal block structure is characteristic for systems
with chiral symmetry. The chiral ensemble was introduced to
provide a background for calculating  the universal part of the
microscopic level density for the Euclidian QCD Dirac operator,
see \cite{Ver} and references therein. Independently and
simultaneously it was realised that the same chiral ensemble is
describing a new group structure associated with scattering in
disordered mesoscopic wires \cite{SN}.

One of the main objects of interest in QCD is the so-called
Euclidean partition function  used to describe a system of quarks
characterized by $n_f$ flavors and quark masses $m_f$ interacting
with the Yang-Mills gauge fields. At the level of Random Matrix
Theory the true partition function is replaced by the matrix
integral:
\begin{equation}\label{partf}
{\cal Z}_{n_f}(\hat{M}_f)=\int {\cal
D}\hat{J}\prod_{k=1}^{n_f}\det\{i\hat{D}+m_f^{(k)}{\bf 1}_{2N}\}
e^{-N \mbox{\small Tr}V\left(\hat{J}^{\dagger}\hat{J}\right)}
\end{equation}
where $V(z)$ is a suitable potential. Here the integration over
complex $\hat{J}$ serves to mimic the functional integral over
gauge field configurations \cite{Ver}. Then the calculation of the
partition function amounts to performing the ensemble average of
the product of characteristic polynomials of $i\hat{D}$ over the
probability density $P(J)\propto e^{-N \mbox{\small
Tr}V\left(\hat{J}^{\dagger}\hat{J}\right)}$ . To model the
simplest case of the sector with zero topological charge the
matrices $\hat{J}$ have to be chosen general $N\times N$ complex,
and the probability distribution can be chosen Gaussian as defined
by the formula:
\begin{equation}\label{ens}
P(\hat{J})d\hat{J}d\hat{J}^{\dag}=\mbox{const}\exp\left(-N\mbox{Tr}
\left(\hat{J}^{\dag}\hat{J}\right)\right)
\prod\limits_{i,j=1}^{N}(dReJ_{ij})(dImJ_{ij})
\end{equation}
 In a more general case of
non-zero topological charge the matrices $\hat{J}$ have to be chosen
rectangular \cite{Ver}, see also\cite{AW}. One may also wish to consider
a more general type of potentials $V$\cite{DN,Ak}. We do not consider
these modifications in the present paper.

The characteristic feature of the chiral ensemble is the presence
of a particular point $\lambda=0$ in the spectrum, also called the
"hard edge" \cite{For}. The eigenvalues of chiral matrices appear
in pairs $\pm\lambda _k\,\,,\,\, k=1,...,N$. Far from the hard
edge the statistics of eigenvalues is practically the same as for
usual GUE matrices without chiral structure, but in the vicinity
of the edge eigenvalues behave very differently.

Define the following characteristic polynomials for chiral
matrices:
\begin{equation}
{\mathcal{Z}}(\hat{J},m)=\det{\left(m{\bf
1}_{2N}+i\hat{D}\right)}=\det\left(
\begin{array}{cc}
  m{\bf 1}_N & i\hat{J} \\
  i\hat{J}^{\dag} & m{\bf 1_N}
\end{array}
\right).
\label{charpol}
\end{equation}

To calculate the eigenvalue density $\rho(\lambda)$
 one can use the trace of the resolvent
for the matrix $-i\hat{D}$ which is given in terms
of the eigenvalues by
\[
R_{\pm}(m)=\mbox{Tr}\frac{1}{m{\bf 1}_{2N}\pm i\hat{D}}=
2m\sum_{k=1}^N\frac{1}{m^2+\lambda^2_k}
\]
The eigenvalue density is then extracted in the usual way as
\[
\rho(\lambda)=\lim_{\epsilon\to 0}
\frac{1}{\pi}\mbox{Re Tr}
\frac{1}{m{\bf 1}_{2N}+i\hat{D}}\left.\right|_{m=\epsilon+i\lambda}
\]
In turn, the trace of the resolvent can be most easily found via
the identity: $ \mbox{Tr}(m_b{\bf 1}_{2N}
+i\hat{D})^{-1}=\frac{\partial}{\partial m_f}
\frac{Z_N(\hat{J},m_f)}{Z_N(\hat{J},m_b)}|_{m_f=m_b}$. We
therefore see that the interest in spectral properties  of chiral
matrices suggests to consider a general correlation function
containing both products and ratios of the characteristic
polynomials:
\begin{equation}\label{rationofpolynomials}
{\mathcal{K}}(\hat{M}_f,\hat{M}_b)=\left\langle\frac{\prod\limits_{k=1}^{n_f}
{\mathcal{Z}}\left(\hat{J},m_f^{(k)}\right)}{\prod\limits_{l=1}^{n_b}
{\mathcal{Z}}\left(\hat{J},m_b^{(l)}\right)}\right\rangle_{J}
\end{equation}
where
$\hat{M}_f=\mbox{diag}\left(m_f^{(1)},\ldots,m_f^{(n_f)}\right)$,
$\hat{M}_b=\mbox{diag}\left(m_b^{(1)},\ldots,m_b^{(n_b)}\right)$
and the average is taken over the ensemble Eq.(\ref{ens}). In
fact, correlation functions of such a type contain the most
 detailed information about the spectra of random matrices and
can be used in various contexts and applications \cite{AS}.

There are several analytical techniques for
 dealing with the correlation function
 $\left\langle\prod_{k=1}^{N_f}{\mathcal{Z}}(\hat{J},m_k)\right\rangle_J$
of only products of characteristic polynomials. The majority of
the results obtained so far resorted to exploitation of the
orthogonal system of Laguerre polynomials, see
\cite{SN,VeZa,For,TrWi,WGW,DN,BH,Ak}. Another important tool in
QCD applications of the chiral ensemble (as well as in studies on
closely related non-Hermitian random matrices \cite{HP}) proved to
be the following generalisation of the Itzykson-Zuber type
integral
\begin{equation}\label{TINTEGRAL1}
\int\exp\left[-\frac{1}{2}
\mbox{Tr}\left(\hat{X}_d\left[\hat{U}\hat{Y}_d\hat{V}^{\dag}+\hat{V}\hat{Y}_d\hat{U}^{\dag}
\right]\right)\right]
d\mu(\hat{U})d\mu(\hat{V})=\mbox{const}\;\frac{\mbox{det}
\left[I_0(x_iy_j)\right]\vert_{1\leq i,j\leq
n}}{\triangle(x_1^2,\ldots ,x_n^2)\;\triangle(y_1^2,\cdots
,y_n^2)}
\end{equation}
Here the integration in (\ref{TINTEGRAL1}) goes over unitary
$n\times n$ matrices $\hat{U},\hat{V}\in {\sf{U(n)}}$ with
$d\mu(\hat{U},\hat{V})$ being the corresponding invariant
measures. The matrices $\hat{X}_d$ and $\hat{Y}_d$ are diagonal:
\begin{equation}
\hat{X}_d=\mbox{diag}\left(x_1,x_2,\ldots ,x_n\right),\;\;
\hat{Y}_d=\mbox{diag}\left(y_1,y_2,\ldots ,y_n\right)
\end{equation}
and $I_0(x)$ denotes the Bessel function of imaginary argument and
$0^{\mbox{th}}$ order. In fact, the integral formula
(\ref{TINTEGRAL1}) was known already to Berezin and Karpelevich
\cite{BK}, and rediscovered very recently by Guhr and Wettig
\cite{guhr} and by Jackson, Sener and Verbaarschot \cite{JSV1}.

A more general correlation function Eq.(\ref{rationofpolynomials})
 can be studied for the particular case of the
 Gaussian measure ( chiral Gaussian Unitary Ensemble, eq.(\ref{ens}))
 by the procedure known
in the literature as the supermatrix
(or "supersymmetry") approach.
The method was pioneered by Efetov \cite{Efetov}
in the theory of disordered systems.
First, one represents each of the characteristic polynomials in the
numerator  as the Gaussian
integral over anticommuting (Grassmann) variables. Similarly,
characteristic polynomials in the denominator are represented
by the Gaussian integrals over usual commuting complex vectors.
This allows to average the resulting expressions in a simple way.
The resulting multiple non-Gaussian integral
is brought to a tractable form by the so-called
Hubbard-Stratonovich transformation. The latter transformation
has, in general,  a quite complicated
analytical structure which mixes commuting and anticommuting
as well as compact and non-compact integration
variables. Such a trick allows one to expose the integration
variables amenable to the saddle-point treatment in the limit of
large matrix size $N$.

 The first applications of the supersymmetry technique to
the chiral case \cite{AST,JSV2} showed that the computation was
quite technically involved even for the lowest correlation
functions. In fact, any explicit calculation beyond the two-point
correlation function of the resolvents seemed to be problematic.

In our paper \cite{II} we demonstrated that an alternative
 method suggested in
\cite{I} enabled us to calculate general spectral correlation functions
 for random Hermitian matrices without chiral structure
 in a systematic and efficient way.
Our starting point was the same as for the supersymmetry approach.
 At the same time, we integrated out
Grassmann variables at a very early stage and avoided the complicated
Hubbard-Stratonovich transformation thus seriously departing from the
general spirit of supersymmetry.

In the present paper we show that the methods developed in
\cite{I,II} can be successfully applied to calculate the general
correlation function of products and ratios of characteristic
polynomials for chiral GUE matrices (for the simplest case
$n_b=n_f=1$ this fact was already demonstrated in \cite{I}).

Our main result is the formula:
\begin{equation}\label{main}
{\mathcal{K}}_N(\hat{X})=\mbox{const}\frac{e^{-\frac{1}{4N}\mbox{Tr}\hat{X}^2_F}}
{\Delta\{\hat{X}^2_B\}\Delta\{\hat{X}^2_F\}}
\det{\left(\int_0^{\infty}dR e^{-NR} R^{N-n_b+i-1}{\cal
J}_k(X^{(k)}\sqrt{R})\right)}_{i,k=1}^n
\end{equation}
valid for any values of the parameters
$N,n_b,n_f,\hat{X}_B=2N\hat{M}_b,\hat{X}_F=2N\hat{M}_f$, provided
$N\ge n_b$. Here we introduced the matrix
$\hat{X}=\mbox{diag}\left(\hat{X}_F,\hat{X}_B\right)$ of the size
$n=n_f+n_b$ and denoted
$$
{\cal J}_k(z)=\left\{\begin{array}{l} I_0(z)\quad, \,1\le k\le n_f\\
K_0(z)\quad, \,n_f+1\le k\le n\end{array}\right.
$$
where $K_0(x)$ denotes the Macdonald function of
$0^{\mbox{th}}$ order.

Quite remarkably, the final expression is represented in
a compact determinantal form reminiscent of that arising from
the orthogonal polynomial method for positive moments.
In a separate publication \cite{III} we show that such a
determinantal form is in no way accidental and is not specific
for the chiral case. In fact, we managed to derive
it in full generality not only for the Gaussian distribution, but
 for arbitrary unitary-invariant potential.
For the present case of chiral GUE one can perform the asymptotic
analysis of Eq.(\ref{main}) in the "chiral limit" $N\to\infty$
with fixed $\hat{X}_B$ and $\hat{X}_F$ and obtain  more compact
determinant expression
 Eq.(\ref{fin}).
This is also a new result going beyond the known asymptotic expressions
for the product of positive \cite{WGW,JSV1,BH}
as well as only negative \cite{I} moments.

The key technical tool for our analysis is the matrix integral:
\begin{equation}\label{TINTEGRAL2}
\int\exp\left[-\frac{1}{2}\mbox{Tr}\left(\hat{X}_d
\left[\hat{T}\hat{Y}_d\hat{T}^{\dag}+\left(\hat{T}^{\dag}\right)^{-1}
\hat{Y}_d\hat{T}^{-1}\right]\right)\right]
d\mu(\hat{T},\hat{T}^{\dagger})=\mbox{const}\;\frac{\mbox{det}\left[K_0(x_iy_j)\right]\vert_{1\leq
i,j\leq n}}{\triangle(x_1^2,\ldots ,x_n^2)\;\triangle(y_1^2,\cdots
,y_n^2)}
\end{equation}
Now $\hat{X}_d$ and $\hat{Y}_d$ are
two positive definite diagonal matrices:
\begin{equation}
\hat{X}_d=\mbox{diag}\left(x_1,x_2,\ldots ,x_n\right)>0,\;\;
\hat{Y}_d=\mbox{diag}\left(y_1,y_2,\ldots ,y_n\right)>0
\end{equation}
The integration in (\ref{TINTEGRAL2}) goes over $\hat{T}\in
{\sf{Gl(n,{\mathcal{C}})/U(1)\times ... \times U(1)}}$, i.e. over
arbitrary $n\times n$ complex matrices with positive diagonal
elements, with $d\mu(\hat{T},\hat{T}^{\dagger})$ being the corresponding invariant
measure.

The integrals (\ref{TINTEGRAL1}) and (\ref{TINTEGRAL2})
 can be looked at as certain matrix
Bessel and Macdonald functions.
 In fact, second integral
(\ref{TINTEGRAL2}) is a very natural non-compact counterpart of the
integral (\ref{TINTEGRAL1}).

The structure of the paper is as follows. First we follow
the (improved) version of the method suggested in \cite{I} and
derive a convenient representation of the
general correlation function in terms of a matrix integral.
Then we integrate out irrelevant degrees of freedom
exploiting Eqs.(\ref{TINTEGRAL1})
and (\ref{TINTEGRAL2}) and reveal a simple determinantal
structure of the resulting expression. As a by-product we
 expose variables amenable
to saddle-point treatment in the limit of large matrix size $N$.
Finally we calculate the resulting correlation functions
in the "chiral scaling" limit $N\to\infty, mN=x<\infty$.
In the Appendices we present a derivation of the Itzykson-Zuber-like
integral (\ref{TINTEGRAL2}) using standard "diffusion equation"
arguments, as well as discuss some technical details of matrix
parametrisations used in our
calculation.

\section{Correlation Functions for Chiral GUE}
As we discussed in the Introduction the starting point of our
method is similar to the standard supersymmetry approach
\cite{Efetov} and is
 based on a simultaneous exploitation of commuting
(or bosonic)  and anticommuting Grassmann (or fermionic)
variables. We introduce $2n_f$ fermionic and $2n_b$ bosonic
vectors, each of them with $N$ components:
\begin{itemize}
  \item ${\bf\chi}_k, {\bf\varphi}_k$ - fermionic vectors,
  $k=1,2,\ldots ,n_f$
  \item ${\bf s}_l,\; {\bf p}_l$ -  bosonic vectors, $l=1,2,\ldots
  ,n_b$
\end{itemize}
These vectors enable us to rewrite the ratio of products of
characteristic polynomials in Eq. (\ref{rationofpolynomials}) as
the following integral:
\begin{equation}\label{ratioasintegral}
\frac{\prod\limits_{k=1}^{n_f}
{\mathcal{Z}}\left(J,m_f^{(k)}\right)}{\prod\limits_{l=1}^{n_b}
{\mathcal{Z}}\left(J,m_b^{(l)}\right)}=
\mbox{const}\;\int{\mathcal{D}}{\mathcal{F}}
{\mathcal{D}}{\mathcal{B}}\;\exp\left\{-{\mathcal{A}}_{\mathcal{F}}(J)-
{\mathcal{A}}_{\mathcal{B}}(J)\right\}
\end{equation}
where we have denoted:
\begin{equation}
{\mathcal{D}}{\mathcal{F}}
\equiv\prod\limits_{k=1}^{n_f}d{\bf{\chi}}_k^{\dag}d{\bf{\chi}}_kd{\bf{\varphi}}^{\dag}_kd{\bf{\varphi}}_k,\;\;
{\mathcal{D}}{\mathcal{B}}
\equiv\prod\limits_{l=1}^{n_b}d{\bf{s}}_l^{\dag}d{\bf{s}}_ld{\bf{p}}^{\dag}_ld{\bf{p}}_l
\end{equation}
and
\begin{equation}
{\mathcal{A}}_{\mathcal{F}}(J)=\sum\limits_{k=1}^{n_f}\left[m_f^{(k)}\left({\bf{\chi}}_k^{\dag}{\bf{\chi}}_k
+{\bf{\varphi}}_k^{\dag}{\bf{\varphi}}_k\right)+
i\;{\bf{\varphi}}_k^{\dag}\hat{J}^{\dag}{\bf{\chi}}_k+
i\;{\bf{\chi}}_k^{\dag}\hat{J}\varphi_k\right]
\end{equation}
\begin{equation}
{\mathcal{A}}_{\mathcal{B}}(J)=\sum\limits_{l=1}^{n_b}
\left[m_b^{(l)}\left({\bf{s}}_l^{\dag}{\bf{s}}_l
+{\bf{p}}_l^{\dag}{\bf{p}}_l\right)+i\;{\bf{p}}_l^{\dag}\hat{J}^{\dag}{\bf{s}}_l+
i\;{\bf{s}}_l^{\dag}\hat{J}{\bf p}_l\right]
\end{equation}
The result of the above manipulation is that we can easily perform the
ensemble average over the matrices $J$ in
Eq.(\ref{rationofpolynomials}) using the identity:
\begin{equation}\label{AV}
\left\langle
e^{-Tr(\hat{J}^{\dagger}\hat{C}+\hat{D}\hat{J})}\right\rangle_J
=e^{\frac{1}{N}Tr{\hat{C}\hat{D}}}
\end{equation}
where in our case:
$$
\hat{C}=i\sum_l {\bf s}_l\otimes{\bf p}_l^{\dagger}-i\sum_k{\bf \varphi}_k
\otimes{\bf {\chi}}^{\dagger}_k\quad,\quad \hat{D}=
i\sum_l {\bf p}_l\otimes{\bf s}_l^{\dagger}-i\sum_k{\bf \chi}_k
\otimes{\bf \varphi}^{\dagger}_k
$$

 After inserting
Eq.(\ref{ratioasintegral}) into Eq. (\ref{rationofpolynomials})
and performing the average in the way described above we find it
convenient to introduce two matrices
$\hat{\mathcal{X}},\hat{\mathcal{Y}} $ of the size $n_f\times n_f$
with entries:
\begin{equation}
{\mathcal{X}}_{k_1k_2}={\bf{\chi}}_{k_1}^{\dag}{\bf{\chi}}_{k_2},\;\;
{\mathcal{Y}}_{k_1k_2}={\bf{\varphi}}_{k_1}^{\dag}{\bf{\varphi}}_{k_2},\;\;
\end{equation}
as well as two  matrices $\hat{Q}_{B1},\hat{Q}_{B2}$ of the size
$n_b\times n_b$ :
\begin{equation}\label{qb}
Q_{B1}^{l_1l_2}={\bf{s}}_{l_1}^{\dag}{\bf{s}}_{l_2},\;\;
Q_{B2}^{l_1l_2}={\bf{p}}_{l_1}^{\dag}{\bf{p}}_{l_2}
\end{equation}
Then the result of the average can be written in a compact form as
\begin{equation}\label{resultofaverage}
{\mathcal{K}}(\hat{M}_f,\hat{M}_b)=\mbox{const}\;\int{\mathcal{D}}
{\mathcal{F}}{\mathcal{D}}{\mathcal{B}}\exp\left\{-{\mathcal{W_{F}}}
-{\mathcal{W_{B}}}-{\mathcal{W_{FB}}}\right\}
\end{equation}
where
\begin{equation}
{\mathcal{W_F}}=\mbox{Tr}
\left\{M_f(\hat{\mathcal{X}}+\hat{\mathcal{Y}})-
\frac{1}{N}\hat{\mathcal{X}}\hat{\mathcal{Y}}\right\}
\end{equation}
\begin{equation}
{\mathcal{W_B}}=\mbox{Tr}\left\{\hat{M}_b(\hat{Q}_{B1}+\hat{Q}_{B2})+
\frac{1}{N}{\hat{Q}_{B1}}{\hat{Q}_{B2}}\right\}
\end{equation}
\begin{equation}
{\mathcal{W_{FB}}}=\frac{1}{N}\sum\limits_{k,l}\left[{\bf{\chi}}_{k}^{\dag}({\bf{s}}_l\otimes
{\bf{p}}_l^{\dag})\phi_k+\phi^{\dag}_{k}({\bf{p}}_l\otimes{\bf{s}}_l^{\dag})\chi_k\right]
\end{equation}
In contrast to the standard supersymmetric approach from now on
we deal with
fermionic and bosonic integration separately. Let us first
integrate out fermionic variables in Eq.(\ref{resultofaverage}).
To remove the non-gaussian terms in the exponent we employ the
simplest version of the
Hubbard-Stratonovich transformation (cf. (\ref{AV})):
\begin{equation}
\exp\left[\frac{1}{N}\mbox{Tr}
\left(\hat{\mathcal{X}}\hat{\mathcal{Y}}\right)\right]=
\mbox{const}\;\int d\hat{Q}_F d\hat{Q}_F^{\dag}
\exp\left\{-N\mbox{Tr}(\hat{Q}_F^{\dag}\hat{Q}_F)-
\mbox{Tr}(\hat{Q}_F\hat{\mathcal{X}})-\mbox{Tr}(\hat{Q}_F^{\dag}
\hat{\mathcal{Y}})\right\}
\end{equation}
where $\hat{Q}_F$ is a $n_f\times n_f$ complex matrix. We insert
the above formula to the integral Eq.(\ref{resultofaverage}) and
change the order of integrations. Then the integration over
fermionic vectors can be done explicitly yielding
 the following representation for the correlation
function:
\begin{equation}\label{ququ}
{\mathcal{K}}(\hat{M}_f,\hat{M}_b)=\int{\mathcal DB}d\hat{Q}_Fd\hat{Q}_F^{\dag}
\,\det{\hat{A}}\,\exp\left\{-{\mathcal W_B}-N\mbox{Tr}(\hat{Q}^{\dag}_F\hat{Q}_F)\right\}
\end{equation}
where the $(2n_f N)\times (2n_f N)$ matrix $\hat{A}$ has the
following structure:
\begin{equation}
\hat{A}=\left(\begin{array}{cc}
  \hat{m}_F\otimes 1_N & \frac{1}{N}\left(1_{n_f}\otimes
\hat{\mathcal B}\right) \\
  \frac{1}{N}\left(1_{n_f}\otimes
\hat{\mathcal B}\right)^{\dag} & \hat{m}^{\dag}_F\otimes 1_N
\end{array}
\right)
\end{equation}
with $\hat{m}_F=\hat{M}_f+\hat{Q}_F,\;\;\mbox{and}
\;\;\hat{\mathcal
B}=\sum\limits_{l=1}^{n_b}{\bf{s}}_l\otimes({\bf{p}}^{\dagger}_l)$.
By simple algebraic manipulations one can show that:
$$
\mbox{det}\left(\begin{array}{cc}
  \hat{a}\otimes 1_N & \left(1_{n_f}\otimes
\hat{b}\right) \\
  \left(1_{n_f}\otimes
\hat{b}\right)^{\dag} & \hat{a}^{\dag}\otimes 1_N
\end{array}
\right)=\mbox{det}\left(\left[\hat{a}^{\dagger}\hat{a}\right]\otimes
1_N -1_{n_f}\otimes\left[ \hat{b}\hat{b}^{\dagger}\right]\right)
$$
for any $n_f\times n_f$ matrix $\hat{a}$ and $N\times N$ matrix
$\hat{b}$. Moreover, it is easy to see that
$$
\mbox{det}\left(\hat{R}_a\otimes 1_N
-1_{n}\otimes\hat{R}_b\right)=\prod_{i=1}^n\mbox{det}
\left(r^{(i)}_a{\bf 1}_N-\hat{R}_b\right)=
\prod_{i=1}^n\prod_{j=1}^N
\left(r^{(i)}_a-r^{(j)}_b\right)
$$
for any two Hermitian matrices $\hat{R}_a$ and $\hat{R}_b$ with
eigenvalues $r^{(1)}_a,...,r^{(n)}_a$ and
$r^{(1)}_b,...,r^{(N)}_b$, respectively. Finally, we notice that
the matrix $\hat{\cal R}_B=\hat{\mathcal B} \hat{\mathcal
B}^{\dagger}$ has obviously rank $n_b$, i.e. $n_b$ nonzero
eigenvalues,
 and verify that $\mbox{Tr}\hat{\cal R}_B^p=\mbox{Tr}\left(
\hat{Q}_{B1}\hat{Q}_{B2}\right)^p$ for any positive integer $p$.
As a consequence, for $n_b\le N$ we have
$$
\mbox{det}\left(r^{(i)}_a{\bf 1}_N-N^{-2}\hat{\cal R}_B\right)
=\left[r^{(i)}_a\right]^{N-n_b} \mbox{det}\left(r^{(i)}_a{\bf
1}_{n_b}-N^{-2}\hat{Q}_{B1}\hat{Q}_{B2}\right)
$$
and immediately obtain the following formula for
the determinant entering the equation Eq.(\ref{ququ}):
\begin{equation}
\det{\hat{A}}=\det\left[ \hat{m}_F
\hat{m}^{\dagger}_F\right]^{N-n_b}
\mbox{det}\left(\hat{m}^{\dagger}_F\hat{m}_F\otimes 1_{n_b} -
\frac{1}{N^2}\left(1_{n_f}\otimes
\left[\hat{Q}_{B1}\hat{Q}_{B2}\right] \right) \right)
\end{equation}

We see now that the integrand depends on vectors ${\bf s}_l,{\bf
p}_l\,,\, l=1,...,n_b$ only via the scalar products ${\bf
s}^{\dagger}_{l1}{\bf s}_{l_2}$ and ${\bf p}^{\dagger}_{l1}{\bf
p}_{l_2}$ forming the matrices $\hat{Q}_{B(1,2)}$, see
Eq.(\ref{qb}). Then one can convert integration over vectors into
integration over the Hermitian matrices
$\hat{Q}_{B1}>0,\hat{Q}_{B_2}>0$ using the integration theorem,
see \cite{I,II}:
$$
\int {\cal DB}\,\, {\cal F}\left(\hat{Q}_{B1},\hat{Q}_{B2}\right)
\propto \int_{\hat{Q}_{B1}>0} d\hat{Q}_{B1}
 \int_{\hat{Q}_{B2}>0} d\hat{Q}_{B2}
\,\, {\cal F}\left(\hat{Q}_{B1},\hat{Q}_{B2}\right)
\left[\det{\left(\hat{Q}_{B1}\right)}
\det{\left(\hat{Q}_{B2}\right)}\right]^{N-n_b}
$$
and to write down the correlation function under consideration
as a matrix integral:
\begin{eqnarray}\label{intrep}
{\mathcal{K}}_N(\hat{X}_F,\hat{X}_B)
&\propto&e^{-\frac{1}{4N}\mbox{Tr}\hat{X}^2_F} \int d\hat{Q}_{F}
\int d\hat{Q}^{\dagger}_{F} \,\,
e^{\frac{1}{2}\mbox{Tr}\hat{X}_F\left(\hat{Q}^{\dagger}_{F}+\hat{Q}_{F}\right)
-N\mbox{Tr}\left(\hat{Q}^{\dagger}_{F}\hat{Q}_{F}\right)}
\left[\det{\left(\hat{Q}^{\dagger}_{F}\hat{Q}_{F}\right)}\right]^{N-n_b}
\\ \nonumber
&\times& \int_{\hat{Q}_{B1}>0}d\hat{Q}_{B1} \int_{\hat{Q}_{B2}>0}
d\hat{Q}_{B2} \,\,
e^{-\frac{1}{2}\mbox{Tr}\hat{X}_B\left(\hat{Q}_{B1}+\hat{Q}_{B2}\right)
-N\mbox{Tr}\left(\hat{Q}_{B1}\hat{Q}_{B2}\right)}
\left[\det{\left(\hat{Q}_{B1}\hat{Q}_{B2}\right)}\right]^{N-n_b}
\\ \nonumber &\times&
\det{\left[\left(\hat{Q}^{\dagger}_{F}\hat{Q}_{F}\right) \otimes
1_{n_b}-1_{n_f}\otimes\left(\hat{Q}_{B1}\hat{Q}_{B2}\right)
\right]}
\end{eqnarray}
where we denoted $\hat{X}_{B}=2N\,\hat{M}_{b}$ and $\hat{X}_{F}=2N\,\hat{M}_{f}$,
 correspondingly.

Next natural step is to use the singular value decomposition
 $\hat{Q}_F=\hat{U}^{\dagger}\,\hat{q}\,\hat{V}$, where $\hat{U},\hat{V}\in
{\sf{U(n_f)}}$ are two different $n_f\times n_f$ unitary matrices,
and $\hat{q}=\mbox{diag}(q_{1},...,q_{n_f})$ are singular values,
i.e. positive square roots of the eigenvalues of
$\hat{Q}_F^{\dagger} \hat{Q}_F>0$.
 The integration measure in the new variables is given by
$d\hat{Q}^{\dagger}_Fd\hat{Q}_F\propto \prod_{k=1}^{n_f}q_k\,dq_k
\Delta^2\{\hat{q}^2\}d\mu(\hat{U})d\mu(\hat{V})$, with
$d\mu(\hat{U},\hat{V})$ being the corresponding Haar's measure on
the group ${\sf{U(n_f)}}$ and
$\Delta\{\hat{q}^2\}=\prod_{k_1<k_2}\left(q^2_{k_1}-q^2_{k_2}\right)$.

Now we have to find an appropriate parameterisation for the pair
of positive definite Hermitian matrices $\hat{Q}_{B1}>0$ and
$\hat{Q}_{B2}>0$ of the size $n_b\times n_b$. One can prove that
such a pair can always be uniquely represented as
\begin{equation}\label{par}
\hat{Q}_{B1}=\left[\hat{T}^{\dagger}\right]^{-1}\hat{P}\left[\hat{T}\right]^{-1}
\quad,\quad \hat{Q}_{B2}=\hat{T}\hat{P}\hat{T}^{\dagger}
\end{equation}
in terms of a positive definite diagonal $\hat{P}>0$ and a general
complex matrix $\hat{T}$ with real positive diagonal entries:
$\hat{T}\in {\sf{GL(n,{\mathcal{C}})/U(1)\times\ldots \times
U(1)}}$. The corresponding integration measure is:
\begin{equation}\label{measure}
d\hat{Q}_{B1}d\hat{Q}_{B_2}\propto \prod_{l=1}^{n_b}p_l dp_l
\prod_{l<m}(p_l^2-p_m^2)^2d\mu(\hat{T},\hat{T}^{\dagger})\quad,\quad
d\mu(\hat{T},\hat{T}^{\dagger}) =d\hat{T} d\hat{T}^{\dagger}
\det{\left[\hat{T}\hat{T}^{\dagger}\right]}^{-n_b+\frac{1}{2}}
\end{equation}
The proof of the above statements is presented in the Appendix A.

The only terms in the integrand Eq.(\ref{intrep}) which depend on
the matrices $\hat{U},\hat{V}$ and $\hat{T},\hat{T}^{\dagger}$ are
$e^{\frac{1}{2}\mbox{Tr}\hat{X}_F\left(\hat{Q}^{\dagger}_{F}+
\hat{Q}_{F}\right)}$ and
$e^{-\frac{1}{2}\mbox{Tr}\hat{X}_B\left(\hat{Q}_{B1}+\hat{Q}_{B2}\right)}$,
respectively. It is immediately evident that the integration over
$d\mu(\hat{U},\hat{V})$ can be performed using the matrix Bessel
function Eq.(\ref{TINTEGRAL1}) whereas its counterpart
Eq.(\ref{TINTEGRAL2}) allows us to integrate out the matrices
$\hat{T}$ and $\hat{T}^{\dagger}$. Moreover, the Vandermonde
determinant factors arising in the denominator after that
integration cancel away one of such factors coming from the
integration measure. As the result, the integrand becomes
proportional to fully antisymmetric Vandermonde factor. This
antisymmetry can be used to show that the integral of the
determinant of Bessel functions is equal to the integral of the
product of the diagonal elements of the corresponding $n_f\times
n_f$ matrix, multiplied with the overall $n_f!$ factor.(This
factor counts the number of terms in the determinant). The same
procedure is applicable to the integral of the $n_b\times n_b$
determinant containing Macdonald functions. Denoting
$\hat{P}^2=\hat{R}_B\,\,,\,\, \hat{q}^2=\hat{R}_F$ with
eigenvalues $R_B^{(l)}$ and $R_F^{(k)}$, respectively, the
resulting expression is given by:
\begin{eqnarray}
\nonumber
{\mathcal{K}}_N(X_F,X_B)\propto\frac{e^{-\frac{1}{4N}\mbox{Tr}\hat{X}^2_F}}
{\Delta\{\hat{X}^2_B\}\Delta\{\hat{X}^2_F\}}
\qquad\qquad\qquad\qquad\qquad
 \nonumber\\
\times\int_{\hat{R}_F>0}d\hat{R}_F \Delta\{\hat{R}_F\}
\left[\det{\hat{R}_F}\right]^{N-n_b}e^{-N\mbox{Tr}\hat{R}_F}
\prod_{k=1}^{n_f}I_0\left(X_{F}^{(k)}\sqrt{R_{F}^{(k)}}\right)
\qquad\qquad
\nonumber\\
\times \int_{\hat{R}_B>0}d\hat{R}_B\Delta\{\hat{R}_B\}
\left[\det{\hat{R}_B}\right]^{N-n_b}
e^{-N\mbox{Tr}\hat{R}_B}\prod_{l=1}^{n_b}K_0\left(X_{B}^{(l)}\sqrt{R_{B}^{(l)}}
\right)
\prod_{k=1}^{n_f}\prod_{l=1}^{n_b}(R_{F}^{(k)}-R_{B}^{(l)})
\end{eqnarray}

We further introduce a matrix
$\hat{R}=\mbox{diag}(\hat{R}_F,\hat{R}_B)$ of the size $n=n_f+n_b$
and another matrix
$\hat{X}=\mbox{diag}\left(\hat{X}_F,\hat{X}_B\right)$ of the same
size. Then
\begin{equation}\label{Kmain}
{\mathcal{K}}_N(\hat{X})=C_N\frac{e^{-\frac{1}{4N}\mbox{Tr}\hat{X}^2_F}}
{\Delta\{\hat{X}^2_B\}\Delta\{\hat{X}^2_F\}}
\int_{\hat{R}>0}d\hat{R}\Delta\{\hat{R}\}\left[\det{\hat{R}}
\right]^{N-n_b}e^{-N\mbox{Tr}\hat{R}} \prod_{k=1}^{n}{\cal
J}_k\left(X^{(k)}\sqrt{R^{(k)}}\right)
\end{equation}
where
$$
{\cal J}_k(z)=\left\{\begin{array}{l} I_0(z)\quad, \,1\le k\le n_f\\
K_0(z)\quad, \,n_f+1\le k\le n\end{array}\right.
$$

Finally, the presence of the Vandermonde determinant in
Eq.(\ref{Kmain}) suggests to rewrite the correlation function in
the form of $n\times n$ determinant. This leads us to the
determinant formula Eq.(\ref{main}) which is the principal result
of the present paper. Our formula generalises  known expressions
for the product of positive moments \cite{WGW}. It is valid for
any values of the parameters $N,n_b,n_f,X_B,X_F$, provided $N\ge
n_b$.

The proportionality constant $C_N$ in the above formulae is fixed
by the obvious normalisation condition
${\mathcal{K}}_N(\hat{X})|_{X_B=X_F}=1$ following from the very
definition (\ref{rationofpolynomials}). In fact, we find it easier
to restore this constant considering the limit $X_B\to\infty,
X_F\to \infty$, when obviously ${\mathcal{K}}_N(\hat{X})\to
(2N)^{2N(n_b-n_f)}
\frac{\prod_{k=1}^{n_f}\left[X_F^{(k)}\right]^{2N}}
{\prod_{l=1}^{n_b}\left[X_B^{(l)}\right]^{2N}}$. On the other
hand, inspecting the integral Eq.(\ref{Kmain}) one notices that in
the discussed limit one can effectively put $R^{(l)}_B\ll
R^{(k)}_F$ in the integrand. Then the integral decouples into the
product of two integrals:
${\mathcal{K}}_N(\hat{X})|_{X_B,X_F\to\infty}=
{\mathcal{K}}_B(\hat{X_B})|_{X_B\to\infty}{\mathcal{K}}_F(\hat{X_F})
|_{X_F\to\infty}$, where
\begin{eqnarray}\label{e1}
{\mathcal{K}}_B(\hat{X_B})= \frac{1}{\Delta\left\{\hat{X}^2_B
\right\}} \int_{\hat{R}_B>0}d\hat{R}_B\Delta\{\hat{R}_B\}
\left[\det{\hat{R}_B}\right]^{N-n_b}
e^{-N\mbox{Tr}\hat{R}_B}\prod_{l=1}^{n_b}K_0
\left(X^{(B)}_l\sqrt{R_{B}^{(l)}}\right)\\
{\mathcal{K}}_F(\hat{X_F})= \frac{e^{-\frac{1}{4N}\mbox{\small
Tr}\hat{X}_F^2}} {\Delta\left\{\hat{X}^2_F \right\}}
\int_{\hat{R}_F>0}d\hat{R}_F\Delta\{\hat{R}_F\}
\left[\det{\hat{R}_F}\right]^{N}
e^{-N\mbox{Tr}\hat{R}_F}\prod_{k=1}^{n_f}I_0
\left(X_{F}^{(k)}\sqrt{R_{F}^{(k)}}\right)\label{e2}
\end{eqnarray}
In the limit of large $X^{(l)}_B\gg N^{1/2}$ one can use the asymptotic
expression for the Macdonald function $K_0(x\gg
1)=\sqrt{\frac{\pi}{2x}} e^{-x}$ in the integrand of (\ref{e1}) and
safely neglect the terms $NR_B^{(l)}$ in the exponent. The resulting
integral is of the form
\begin{equation}
\int_0^{\infty}\ldots \int_0^{\infty}
\Delta\{R_1,...,R_n\}\prod_{j=1}^n R_j^{a}e^{-X_j\sqrt{R_j}}dR_j
=2^n\frac{\Delta\{X^2_1,...,X^2_n\}}
{\prod_{j=1}^n X_j^{2(a+n)}}
\prod_{j=1}^n\Gamma\left[2(j+a)\right]
\end{equation}
In our case $a=N-n_b-1/4$ which yields:
\begin{equation}\label{e3}
{\mathcal{K}}_B(\hat{X_B})|_{X_B\to\infty}=(2\pi)^{n_b/2}
\prod_{l=1}^{n_b}\left[X_B^{(l)}\right]^{-2N}
\prod_{l=1}^{n_b}\Gamma\left[ 2(N-n_b+l)-1/2\right]
\end{equation}
For the second integral we should use the asymptotic expression for the
modified Bessel function $I_0(x\gg
1)=\sqrt{\frac{1}{2\pi x}} e^{x}$. Here, however, we can not
neglect the terms $NR_F^{(k)}$ in the exponent since it ensures the
convergence of the integral. In fact, it is easy to see that for
$X_F^{(k)}\gg 2N$ the integral is dominated by small vicinity of
the value $R_F^{(k)}=X_F^{(k)}/2N$. Taking into account small
Gaussian fluctuations around those points we find that
\begin{equation}\label{e4}
{\mathcal{K}}_F(\hat{X_F})|_{X_F\to\infty}=
\prod_{k=1}^{n_f}\left[X_F^{(k)}\right]^{2N}
\frac{1}{2^{2(Nn_f+n_f^2-n_f)}N^{2Nn_f+n_f^2}}
\end{equation}
The equations (\ref{e3})-(\ref{e4}) yield the following value for the
normalisation constant:
$$C_N=\frac{2^{2(Nn_b+n_f^2-n_f)}N^{n_f^2+2Nn_b}}{(2\pi)^{n_b/2}
\prod\limits_{l=i}^{n_b}\Gamma\left[ 2(N-n_b+l)-1/2\right]}$$ As
usual, we would like to extract the leading behavior of the above
correlation function in the scaling limit $N\to \infty$. Here we
are mostly interested in the so-called "chiral scaling"
${\mathcal{K}}^{chir}(\hat{X})=\lim_{N\to\infty}{\mathcal{K}}_N(\hat{X})$
which amounts to keeping the parameters $\hat{X}$ fixed when
treating the corresponding integrals by the saddle-point method.
The saddle-points are extremals of the function ${\cal
L}(R_k)=R_k-\ln{R_k},\,\, k=1,...,n$ in the domain $R_k>0$, which
are given by $R_k=1$. To pick up a nonvanishing contribution one
has to take accurately into account Gaussian fluctuations around
the saddle point. It is done in the most straightforward way by
exploiting the expression (\ref{Kmain}) and putting there
$R_k=1+\xi_k$, with $-\infty<\xi_k<\infty$ parameterizing the
fluctuations. Introducing $\Xi=(\xi_1,...,\xi_n)$ we have:
\begin{eqnarray}\label{888}
{\mathcal{K}}^{chir}(\hat{X})&\propto&\frac{1}
{\Delta\{\hat{X}^2_B\}\Delta\{\hat{X}^2_F\}} \int
d\Xi\Delta\{\Xi\}e^{-2N\mbox{Tr}\Xi^2}
\prod_{k=1}^{n}{\cal J}_k\left(1+\xi_k)\right)\\
\nonumber &=&\frac{1}
{\Delta\{\hat{X}^2_B\}\Delta\{\hat{X}^2_F\}}\lim_{N\to\infty}
\mbox{det}\left[\int d\xi e^{-2N\xi^2}\xi^{l-1}{\cal J}_k(1+\xi)
\right]_{l,k=1}^n
\end{eqnarray}
Evaluating the limit we finally find:
\begin{eqnarray}\label{fin}
{\mathcal{K}}^{chir}(\hat{X})\propto \frac{1}
{\Delta\{\hat{X}^2_B\}\Delta\{\hat{X}^2_F\}}\times\qquad\qquad\qquad\qquad \qquad\nonumber\\
\mbox{det} \left[
\begin{array}{cccc}
I_0\left[X_{F}^{(1)}\right] &
X_F^{(1)}I_0^{(1)}\left[X_{F}^{(1)}\right] 
& \ldots  & (X_F^{(1)})^{n-1}I_0^{(n-1)}\left[X_{F}^{(n_f)}\right]  \\
  \ldots  & \ldots & \ldots & \ldots \\
  I_0\left[X_{F}^{(n_f)}\right] & X_F^{(n_f)}I_0^{(1)}
\left[X_{F}^{(n_f)}\right] & \ldots & (X_F^{(n_f)})^{n-1}I_0^{(n-1)}\left[X_{F}^{(n_f)}\right] \\
K_0\left[X_{B}^{(1)}\right] &
X_{B}^{(1)}K_0^{(1)}\left[X_{B}^{(1)}\right] 
 & \ldots & (X_{B}^{(1)})^{n-1}K_0^{(n-1)}\left[X_{B}^{(1)}\right]  \\
  \ldots & \ldots & \ldots & \ldots \\
  K_0\left[X_{B}^{(n_b)}\right] & X_B^{(n_b)}K_0^{(1)}
\left[X_{B}^{(n_b)}\right] & \ldots &
  (X_B^{(n_b)})^{n-1}K_0^{(n-1)}\left[X_{B}^{(n_b)}\right]
\end{array}
\right]
\end{eqnarray}
where
 $I_0^{(l)}(z)$ and $K_0^{(l)}(z)$  stand for $l$-th derivatives
of the Bessel and Macdonald functions.

 Such an
expression provides us with the most general correlation function
for chiral GUE and generalises earlier results known for
$n_f>0\,,\,n_b=0$ \cite{JSV1,WGW}, $n_f=0\,,\,n_b>0$ \cite{I}, and
$n_f=n_b=(1,2)$\cite{AST}. For example, for $n_f=n_b=1$ the
equation (\ref{fin}) amounts to
\[
{\cal K}^{chir}(\hat{X}_F,\hat{X}_B)\propto
\left[X_FI_1(X_F)K_0(X_B)+X_BI_0(X_F)K_1(X_B)\right]
\]
and is simply related to the formula found in \cite{AST}.

\subsection{Conclusion}

We have demonstrated that the general spectral correlation
functions containing both products and ratios of the
characteristic polynomials of chiral GUE matrices can be
calculated in a closed form. Our method amounted to representing
the characteristic polynomials in terms of the Gaussian integrals
and exploiting the Itzykson-Zuber type integration formulae 
Eqs.(\ref{TINTEGRAL1},\ref{TINTEGRAL2}), both
over compact and non-compact domains. The results were shown to
have an attractively simple determinantal structure reminiscent of
that arising in the method of orthogonal polynomials. In a
separate publication \cite{III} we demonstrate that indeed there
exists a way to calculate
 the general correlation function by a method resorting
to the orthogonal polynomial
technique. This way makes clear that the determinantal structure
observed in the present paper has, in fact, a very profound origin
and allows one to extend our results to other unitary-invariant
ensembles, with or without chiral structure.
 As interesting prospects for future research we would like
to mention a challenging problems of extending the suggested methods to
other symmetry classes (see \cite{Ak,VK}), as well as to ensembles of non-Hermitian
random matrices. The latter ar important for QCD applications
(see e.g. \cite{Ak1} and references therein) and
 for the problems of quantum
chaotic scattering \cite{FK}.

\subsection{Acknowledgement}
The first author (YVF) is grateful to Jac Verbaarschot
for triggering his interest in the problem and useful correspondence.
Important comments by Thomas Guhr and Tilo Wettig as well as useful
communications with Gernot Akemann are much appreciated.
This research was supported by EPSRC grant GR/13838/01 "Random matrices
close to unitary or Hermitian."

\section{Appendix A. A parametrisation for the pair of
positive definite matrices $Q_{B1}$ and $Q_{B2}$}

Given two positive definite complex Hermitian matrices $Q_{B1}>0$
and $Q_{B2}>0$ of the size $n_b\times n_b$ let us introduce
$\tilde{Q}_{B1}=\left[Q_{B1}\right]^{-1}>0$, so that
$$
dQ_{B1}dQ_{B2}= d\tilde{Q}_{B1}dQ_{B2}\left(\det{\tilde{Q}_{B1}}
\right)^{-2n_b}
$$

Now we use that  any pair of complex Hermitian $Q_{B1}>0$ and
$\tilde{Q_{B2}}>0$ can be parameterized as (see, for example
\cite{Lanc})
$$
\tilde{Q}_{B1}=T\,T^{\dagger}\quad,\quad Q_{B_2}=T\,P\,T^{\dagger}
$$
where the matrix $P$ is a diagonal positive definite:
$P=\mbox{diag}\left(p_1,...,p_{n_b}\right)>0$ and the matrix $T$
is general complex, with the only restriction that its diagonal
entries are real and positive: $T\in
{\sf{Gl(n_b,{\mathcal{C}})/U(1)\times... \times U(1)}}$. The
latter condition is necessary to ensure one-to-one correspondence
between the parametrisations. Our goal is to calculate the related
Jacobian.

For this goal we use the following relation between the matrix differentials:
$$
d\tilde{Q}_{B1}=T\left[\delta{T}+\delta{T}^{\dagger}\right]T^{\dagger}
$$
where we denoted $\delta{T}=T^{-1}dT$ and $\delta{T}^{\dagger}=dT^{\dagger}
\left[T^{\dagger}\right]^{-1}$. Analogously
$$
dQ_{B2}=T\left[\delta{T}P+dP+P\delta{T}^{\dagger}\right]T^{\dagger}
$$
Further denoting $\delta Q_{B1}=T^{-1}d\tilde{Q}_{B1}
\left[T^{\dagger}\right]^{-1},\,\, \delta Q_{B2}=T^{-1}dQ_{B2}
\left[T^{\dagger}\right]^{-1}$ we first calculate the Jacobian of
the transformation from the set of variables $\left[\delta
Q_{B1}\right]_{ii},\left[\delta Q_{B2}\right]_{ii}, \left[\delta
Q_{B1}\right]_{i<j},\left[\delta Q_{B1}^*\right]_{i<j}
\left[\delta Q_{B2}\right]_{i<j},\left[\delta
Q_{B2}^*\right]_{i<j}$ to a new set of independent variables
$dp_i,(\delta T)_{ii},(\delta T)_{i<j}, (\delta T)^*_{i<j},(\delta
T)_{i<j},(\delta T)^*_{i<j}$ where $1<i,j<n_b$ and $*$ stands for
the complex conjugation. The calculation is simple and yields the
factor that we symbolically write as
$$
\frac{\delta(Q_{B1},Q_{B2})}{\delta(P,T,T^{\dagger})}
\propto\prod_{i<j}(p_i-p_j)^2
$$
We need also the following intermediate Jacobians:
$$
\frac{d(\tilde{Q}_{B1},Q_{B2})}{\delta(Q_{B1},Q_{B2})}=
\det{(TT^{\dagger})}^{2n_b}
$$

$$
\frac{\delta(T,T^{\dagger})}{d(T,T^{\dagger})}
=\det{\left(TT^{\dagger}\right)}^{-(n_b-1/2)}
$$
Then all the factors taken together yield the full Jacobian:
\begin{eqnarray}
&&\frac{d(\tilde{Q}_{B1},Q_{B2})}{d(P,T,T^{\dagger})}
=\frac{\delta(Q_{B1},Q_{B2})}{\delta(P,T,T^{\dagger})} \times
\frac{d(\tilde{Q}_{B1},Q_{B2})}{\delta(Q_{B1},Q_{B2})}
\times\frac{\delta(T,T^{\dagger})}{d(T,T^{\dagger})}\\ \nonumber
&\propto&\prod_{i<j}(p_i-p_j)^2\det{\left(TT^{\dagger}\right)}^{(n_b+1/2)}
\end{eqnarray}
To arrive to the parametrisation Eq.(\ref{par}) used in the main
text of the paper we change $P\to P^2$ so that $dP\to 2^{n_b}
dP\det{P}$ in the measure and then change $T\to T\,P^{-1/2}$ so
that $T^{\dagger}\to P^{-1/2} T^{\dagger}$. Then
$Q_{B1}=\tilde{Q}^{-1}_{B1}=
\left[TP^{-1}T^{\dagger}\right]^{-1}=\left[T^{\dagger}\right]^{-1}
P T^{-1}$ as required and the matrix $Q_{B2}$ stays equal to
$TPT^{\dagger}$.
 The transformation amounts to change in the measure
$dTdT^{\dagger}\to \det{P}^{-(n_b-1/2)} dTdT^{\dagger}$.

Finally, taking into account the factor
$$
\det{\left[\tilde{Q}_{B1}\right]}^{-2n_b}
=\det{P}^{-2n_b}\det{\left(TT^{\dagger}\right)}^{2n_b}
$$
we arrive to the expression for the measure given in
Eq.(\ref{measure}). Let us note that $d\mu(T,T^{\dagger})$ is
exactly the invariant measure on the manifold
$\sf{Gl(n,{\mathcal{C}})/U(1)\times ... \times U(1)}$.

\section*{Appendix B. Matrix Macdonald functions
associated with integrals over complex matrices} The matrix Bessel
functions that correspond to Itzykson-Zuber-like integrals over
unitary matrices were considered in details by Guhr and Wettig,
Guhr and Kohler \cite{guhr}. Below we consider the matrix
Macdonald functions  that are associated with integrals over
arbitrary complex matrices with real positive diagonal elements.

 Let $X_d$ and $Y_d$ be
two diagonal matrices with real positive elements,
\begin{equation}
X_d=\mbox{diag}\left(x_1,x_2,\ldots ,x_n\right),\;\;
Y_d=\mbox{diag}\left(y_1,y_2,\ldots ,y_n\right)
\end{equation}
 Our aim is to show that the integral:
\begin{equation}\label{TINTEGRAL}
\Phi(X_d,Y_d)=\int\exp\left[-\frac{1}{2}\mbox{Tr}\left(X_d\left[TY_dT^{\dag}+\left(T^{\dag}\right)^{-1}Y_dT^{-1}\right]\right)\right]
d\mu(T)
\end{equation}
(where $T$ are arbitrary $n\times n$ complex matrices) can be
considered as the matrix Macdonald function. We are going to
demonstrate the following expression for the function
$\Phi(X_d,Y_d)$:
\begin{equation}\label{MAINFORMULA}
\Phi(X_d,Y_d)=\mbox{const}\;\frac{\mbox{det}\left[K_0(x_iy_j)\right]\vert_{1\leq
i,j\leq n}}{\triangle(x_1^2,\ldots ,x_n^2)\;\triangle(y_1^2,\cdots
,y_n^2)}
\end{equation}
where $K_0(x)$ is the Bessel function of an imaginary argument of
$0^{\mbox{th}}$ order.

Given three $n\times n$ complex matrices $X, A, B$ we consider
the Laplace operator $D_{X}$ acting  on complex
matrices as:
\begin{equation}
D_{X}=\sum\limits_{i\leq i,j\leq
n}\left(\frac{\partial^2}{\partial(\mbox{Re}X_{ij})^2}+\frac{\partial^2}{\partial(\mbox{Im}X_{ij})^2}\right)
\end{equation}
Then we  construct a function $W(X,A,B)$ with the property
\begin{equation}\label{laplacematrixequation1}
D_{X}W(X,A,B)=\mbox{Tr}(AB)W(X,A,B)
\end{equation}
In particular, the following function
\begin{equation}
W(X,A,B)=\exp\left[-\frac{1}{2}\mbox{Tr}\left(XA+BX^{\dag}\right)\right]
\end{equation}
satisfies the Eq.(\ref{laplacematrixequation1}) as can be
checked by direct calculations. Let us put specifically $A=TY_dT^{\dag}$ and
$B=(T^{\dag})^{-1}Y_dT^{-1}$ in Eq.(\ref{laplacematrixequation1}).
We obtain:
\begin{eqnarray}
D_{X}\exp\left[-\frac{1}{2}\mbox{Tr}\left(X
\left[TY_dT^{\dag}+(T^{\dag})^{-1}Y_dT^{-1}\right]\right)\right]=\nonumber\;\;\;\;\;\;\;\;\;\;\;\\
\mbox{Tr}(Y_d^2)\exp\left[-\frac{1}{2}\mbox{Tr}\left(X\left[TY_dT^{\dag}+(T^{\dag})^{-1}Y_dT^{-1}\right]\right)\right]
\end{eqnarray}
As soon as the integration over complex matrices $T$ commutes
with the Laplace operator $D_{X}$ we conclude that the matrix
function $\Phi(X_d,Y_d)$ defined by the integral Eq.(\ref{TINTEGRAL})
 satisfies the following differential equation:
\begin{equation}\label{MATRIXBESSEL}
D_{X}\Phi(X_d,Y_d)=\mbox{Tr}(Y_d^2)\Phi(X_d,Y_d)
\end{equation}
To derive the explicit formula (Eq.(\ref{TINTEGRAL})) for the
matrix function $\Phi(X_d,Y_d)$ we apply the method proposed by
Guhr and Wettig \cite{guhr}. Using the singular value decomposition
for an arbitrary complex matrix $X$:
\begin{equation}
X=U^{\dag}X_dV,\; U^{\dag}U=1_n,\; V^{\dag}V=1_n
\end{equation}
the radial part $D_{X_d}$ of the Laplace operator $D_{X}$ must have the
following expression:
\begin{equation}
D_{X_d}=\frac{1}{J(x)}\sum\limits_{i=1}^{n}\partial_iJ(x)\partial_i,\;J(x)=\triangle^2(x_1^2,\ldots
,x_n^2)\prod\limits_{i=1}^{n}x_i
\end{equation}
Guhr and Wettig noted \cite{guhr} that the radial
part $D_{X_d}$ of the Laplace operator $D_{X}$ is separable. It
means that for an arbitrary function $f(x_1,x_2,\ldots ,x_n)$ the
following identity holds:
\begin{equation}\label{separability}
D_{X_d}\frac{f(x_1,\ldots ,x_n)}{\triangle(x_1^2,\ldots
,x_n^2)}=\frac{1}{\triangle(x_1^2,\ldots
,x_n^2)}\sum\limits_{k=1}^{n}\left(\frac{\partial^2}{\partial
x^2_k}+\frac{1}{x_k}\frac{\partial}{\partial
x_k}\right)\frac{f(x_1,\ldots ,x_n)}{\triangle(x_1^2,\ldots
,x_n^2)}
\end{equation}
The separability of the operator $D_{X_d}$ enables us to solve the
differential equation (Eq.\ref{MATRIXBESSEL}) and to prove formula
(\ref{TINTEGRAL}). We use the following ansatz for the solution
$\Phi(X_d,Y_d)$ of differential equation (\ref{MATRIXBESSEL}):
\begin{equation}\label{PHIANZATZ}
\Phi(X_d,Y_d)=\frac{\Psi(x_1,\ldots ,x_n;y_1,\ldots
,y_n)}{\triangle(x_1^2,\ldots ,x_n^2)\triangle(y_1^2,\ldots
,y_n^2)}
\end{equation}
We insert the above ansatz to the differential equation
(Eq.\ref{MATRIXBESSEL}). Applying the property
(Eq.\ref{separability}) of the radial part $D_{X_d}$ of the
Laplace operator $D_{X}$ we find:
\begin{equation}\label{PsiSumDif}
\sum\limits_{k=1}^{n}\left(\partial_k^2+\frac{1}{x_k}\partial_k\right)\Psi(x_1,\ldots
,x_n;y_1,\ldots ,y_n)=\left(y_1^2+y_2^2+\ldots
+y_n^2\right)\Psi(x_1,\ldots ,x_n;y_1,\ldots ,y_n)
\end{equation}
The form of the above differential equation suggests to look for a
solution  as a sum over permutations of the set $(1,2,\ldots ,n)$.
Each term in the sum can be taken as a product of $n$ multipliers
and each multiplier is represented by a (kernel) function taken at
different $x_i$ and $y_j$, i.e.
\begin{equation}\label{Psisum}
\Psi(x_1,\ldots ,x_n;y_1,\ldots ,y_n)= \sum\limits_{\sigma \in
\;{\sf{S_n}}}C_{\sigma}\;\phi(x_1,y_{\sigma(1)})\;\phi(x_2,y_{\sigma(2)})\ldots
\phi(x_n,y_{\sigma(n)})
\end{equation}
where $C_{\sigma}$ is some constant dependent on a particular
permutation $\sigma$. To determine
$C_{\sigma}$ we explore the symmetry properties of
the function $\Psi(x_1,\ldots ,x_n;y_1,\ldots ,y_n)$. The function $\Phi(X_d,Y_d)$ is
symmetric under permutations inside the set $\{y_1,y_2,\ldots
,y_n\}$ as can be seen from the
integral representation
(Eq.\ref{TINTEGRAL}). Indeed, the result of any such permutation can be
represented by a permutation matrix $P$ as
\begin{equation}
Y_d\rightarrow Y_d^{P}=PY_dP^{\dag}
\end{equation}
Let us replace the matrix $Y_d$ by $Y_d^{P}$ in the integral
(Eq.\ref{TINTEGRAL}). Changing the variables of integration,
$T_1=TP$, we see that $\Phi(X_d,PY_dP^{\dag})=\Phi(X_d,Y_d)$, i.e.
the integral (Eq.\ref{TINTEGRAL}) is symmetric under permutations
inside the set $\{y_1,y_2,\ldots ,y_n\}$. As soon as
$\Phi(X_d,Y_d)$ and $\Psi(x_1,\ldots ,x_n,y_1,\ldots ,y_n)$ are
related by Eq.(\ref{PHIANZATZ}) the function $\Psi(x_1,\ldots
,x_n;y_1,\ldots ,y_n)$ is antisymmetric under those permutations.
It gives the coefficient $C_{\sigma}$ in the formula
(Eq.\ref{Psisum}):
\begin{eqnarray}
C_{\sigma}=(-1)^{\nu_{\sigma}}
\end{eqnarray}
where $\nu_{\sigma}=1$ for odd permutations and $\nu_{\sigma}=0$
for even permutations.   We then obtain the
following expression for the function $\Psi(x_1,\ldots
,x_n;y_1,\ldots ,y_n)$ from the Eq.(\ref{Psisum}):
\begin{equation}\label{PsiDet}
\Psi(x_1,\ldots ,x_n;y_1,\ldots
,y_n)=\det\left(\phi(x_i,y_j)\right)|_{1\leq i,j\leq n}
\end{equation}
We insert the expression Eq.(\ref{Psisum}) for $\Psi(x_1,\ldots ,x_n;y_1,\ldots
,y_n)$ to the differential equation
(Eq.\ref{PsiSumDif}). As a result we find that
 the kernel $\phi(x,y)$ satisfies the differential
equation:
\begin{equation}
\left[\partial_x^2+\frac{1}{x}\partial_x\right]\phi(x,y)=y^2\phi(x,y)
\end{equation}
Both Bessel functions of an imaginary argument $I_0(xy)$ and the Macdonald
 functions $K_0(xy)$
 satisfy the above differential equation. However
the integral representation (Eq.\ref{TINTEGRAL}) dictates the choice
for the kernel function $\phi(x,y)$.
 Indeed, we note that for $n=1$ the integral (Eq.\ref{TINTEGRAL})
is proportional to the kernel function $\phi(x,y)$. This integral
representation coincides with that for the Macdonald function $K_0(xy)$
and we conclude that $\phi(x,y)=\mbox{const}\;K_0(xy)$. Then equations
(\ref{PHIANZATZ}) and
 (\ref{PsiDet}) yield the desired formula
 (Eq.\ref{MAINFORMULA}).


\end{document}